\documentclass[12pt]{article}
\usepackage{latexsym,graphicx}
\usepackage{amsmath}
\usepackage{amsfonts}
\usepackage{amssymb}
\usepackage{graphicx}
\usepackage{float}
\usepackage[left=2.5cm,top=2.5cm,right=2.5cm,bottom=1.5cm]{geometry}

\begin{document}
\begin{center}
\large{\bf{Anisotropic bulk viscous string cosmological models of the Universe under a time-dependent deceleration parameter }} \\
\vspace{10mm}
\normalsize{ARCHANA DIXIT$^1$, RASHID ZIA$^2$, ANIRUDH PRADHAN$^3$}\\
\vspace{5mm}
\normalsize{$^{1,2,3}$Department of Mathematics, Institute of Applied Sciences \& Humanities, GLA University,\\ 
	Mathura -281 406, Uttar Pradesh, India \\
\vspace{2mm}
	$^1$E-mail:archana.dixit@gla.ac.in \\
\vspace{2mm}
	$^2$E-mail:rashidzya@gmail.com\\
\vspace{2mm}
$^3$Corresponding Author E-mail:pradhan.anirudh@gmail.com }\\
\end{center}
\begin{abstract}
 We investigate a new class of LRS Bianchi type-II cosmological models by revisiting in the paper of Mishra {\it et al} (2013)
 by considering a new deceleration parameter (DP) depending on the time in string cosmology for the modified gravity theory suggested by
 S$\acute{a}$ez \& Ballester (1986). We have considered the energy-momentum tensor proposed by Leteliar (1983) for bulk viscous
and perfect fluid under some assumptions. To make our models consistent with recent astronomical observations, we
 have used scale factor (Sharma {\it et al} 2018; Garg {\it et al} 2019) $ a(t)=\exp{[\frac{1}{\beta}\sqrt{2 \beta t + k}]}$, 
 where $\beta $ and $k$ are positive constants and it provides a time-varying DP. By using the recent constraints ($H_{0}=73.8$,
 and $q_{0} = -0.54$) from SN Ia data in combination with BAO and CMB observations (Giostri {\it et al}, arXiv:1203.3213v2[astro-ph.CO]),
 we affirm $\beta = 0.0062$ and $k = 0.000016$. For these constraints, we have substantiated a new class of cosmological transit models for
which the expansion takes place from early decelerated phase to the current accelerated phase. Also, we have studied some physical,
 kinematic and geometric behavior of the models, and have found them consistent with observations and well established theoretical results
. We have also compared our present results with those of Mishra {\it et al} (2013) and observed that the results in this paper
 are much better, stable under perturbation and in good agreement with cosmological reflections. 
\end{abstract}

\smallskip
{\it Pacs No.:} 04.20.-q, 04.50.Kd, 98.80.Es 


\section{Introduction}
The current astronomical reflexions, modern experimental data from SNe Ia \cite{ref1}$-$\cite{ref6}; CMBR \cite{ref7,ref8}; 
WMAP \cite{ref9}$-$\cite{ref12} have established two main characteristics of the universe: 
(a) the existence of the anisotropic universe at the early stage of the evolution, which in due course of time
attains isotropy, and (b) the current universe is not only expanding but also the rate of expansion is increasing
(i.e. accelerating universe). The SNe Ia measurements indicate a universe which undergoes through a transition
from past decelerating to present accelerating expansion. So, it is a challenge for theoreticians to provide satisfactory
theoretical support to these observations.\\

Friedmann-Robertson-Walker (FRW) spacetime describes spatially homogenized and isotropic universes, which can be appropriate for
the contemporary universe, however since they have higher symmetries, thus it doesn't provide a correct matter description
of the early universe and presents the poor approximations for an  early universe. Therefore, those models
are additional applications for the outline of the whole evolution of the universe, that have an anisotropic nature
in early time and approaches to isotropy at late times. Bianchi space-times offer a decent framework for this.
Out of all, Bianchi type-II (B-type-II) frame of reference plays a very important affirm in making models for the 
measurements of flourish of the universe throughout its early phase. Moreover, B-type-II line-element yields an 
anisotropic spatial curvature. Recently, Asseo and Sol \cite{ref13} and Roy \& Banerjee \cite{ref14} stressed the 
importance of B-type-II and proscribed LRS cosmological model. Kumar and Akarsu \cite{ref15} mentioned B-type-II 
universe with anisotropic dark energy and perfect fluid. Wang \cite{ref16}$-$\cite{ref18} has investigated the models 
of Letelier-type within the theoretical account of LRS B-type-II. In the context of massive string, Pradhan et al. 
\cite{ref19} have analyzed LRS Bianchi type-II spacetime. B-type-II frame of reference is employed to analyze dark 
energy models within the new role of time-dependent DP by Maurya {\it et al} \cite{ref19}. Within the present study, we 
tend to look into LRS B-type-II string models of the universe for perfect and viscous fluid beneath three conditions. \\

Next, although Einstein's general theory of gravity (GR) explains a large number of the astrophysical phenomenon, it fails
to describe some, for instance, the expanding and late time accelerated the expansion of the universe. To deal with these,
many alternative theories are proposed, out of which, Brans \& Dicke \cite{ref21} and S$\acute{a}$ez \& Ballester \cite{ref22} 
scalar-tensor theories are of significant involvement. In the present paper, we have studied the S$\acute{a}$ez-Ballester 
modified theory of gravity. In this theory, Einstein's field equations have been modified by incorporating a dimensionless 
scalar field $\phi$ coupled with the metric $g_{ij}$ in a simple manner. This modification satisfactorily describes the weak 
field in which an accelerated expansion regime appears. This theory also advises an answer to the question of disappeared matter 
in a non-flat FRW universe. \\

In recent years, string cosmology is gaining significant interest. Cosmic strings are topologically stable objects, which could
be shaped throughout a phase transition within the early universe. Cosmic strings make for a significant role to
study in the early universe. It is assumed that cosmic strings bring about to density perturbations, that cause the formation of
galaxies or cluster of galaxies \cite{ref23}. One more necessary feature of the string is that the string tension provides
rise to an efficient anisotropic pressure. Also, the stress-energy of the string coupled with the gravitational field is also
used to elucidate several alternative cosmological phenomena. The pioneer works in string theory were done by many authors 
\cite{ref24,ref25}. LRS B-type-II cosmological models have been discussed \cite{ref26}$-$\cite{ref29} in different context.
Recently, Pradhan {\it et al} \cite{ref30} have looked into string models of accelerated expansion in $f(R,T)$-gravity with a magnetic field. \\

Also, the dissipation effect together with bulk viscosity presents another model of dark energy. Relaxation processes 
related to bulk viscosity effectively reduce the pressure in an expanding system, in comparison the worth prescribed by 
the equation of state $p=\omega \rho$. The effective pressure becomes negative for a sufficiently large viscosity 
that could imitate a dark energy behavior. The idea that the bulk viscosity drives the acceleration of the universe is 
mentioned in \cite{ref31,ref32}.\\ 

In recent years, many researchers \cite{ref33}$-$\cite{ref38} and references therein have investigated the cosmological 
universes in Saez-Ballester modified gravity theory in various contexts. Under above-discussed perspective, the 
S$\acute{a}$ez \& Ballester field equations have been solved in an LRS B-type-II space-time in the presence of a cloud 
of massive string and bulk viscous fluid, under some physically and geometrically viable assumptions. In the present paper, 
we are revisiting the solutions obtained by Mishra {\it et al} \cite{ref39}, by assuming a scale factor 
$a(t)=exp{[\frac{1}{\beta}\sqrt{2\beta t +k}]}$ which resulting into a time-dependent DP having a transition from the 
decelerating universe to presently accelerating universe. \\

The plan of the manuscript is the following. Section $2$ contains definitions and theoretical calculations. Subsec. $2.1$ deals with 
the metric and field equations are mentioned. Subsec. $2.2$ deals with assumptions and under these assumptions, the solution of 
the field equations are found. In Section $3$, we have derived the solutions of field equations for three different cases, 
Subsections $3.1$, $3.2$ \& $3.3$. Results and discussions are given in Sec. $4$. Stability of corresponding solutions is analyzed 
in Sect. $5$. Finally, conclusions are summarized in Sec. $6$.

\section{Definitions and Theoretical Calculations}

\subsection{Metric and Field equations}

We consider an LRS B-type-II space-time \cite{ref39}:
\begin{equation}
	\label{1}
	ds^{2}=-dt^{2}+X^{2}dx^{2}+Y^{2}dy^{2}+2 X^{2}xdydz+(Y^{2}x^{2}+X^{2})dz^{2}
\end{equation}
where $ X=X(t) $, $Y=Y(t)$.\\

The field equation (in gravitational units $ 8\pi G=1$) proposed by S$\acute{a}$ez \& Ballester \cite{ref22}:

\begin{equation}
	\label{2}
	G_{ij}-\omega\phi^{r}(\phi_{,i}\phi_{,j}-\frac{1}{2}g_{ij}\phi_{,k}{\phi^{,k}})=-T_{ij}.
\end{equation}
Here $G_{ij}=R_{ij}-\frac{1}{2}Rg_{ij}$ and $T_{ij}$ stands for the energy-momentum tensor and $\phi$ for the 
scalar field satisfying the equation
\begin{equation}
	\label{3}
	r\phi^{r-1}\phi_{,k}\phi^{,k}+2\phi^{r}\phi_{;i}^{,i}=0
\end{equation}

Here $\omega$ and $r$ stand for a dimensionless coupling and arbitrary constant respectively. A comma denotes the partial derivative 
whereas a semi-colon denotes partial covariant differentiation w. r. to  $t$.\\

$T_{ij}$, for a cloud of massive string \& bulk viscous fluid, reads:

\begin{equation}
	\label{4}
	T_{ij}=\overline p g_{ij}-\lambda x_{i}x_{j}+(\rho +\overline p) v_{i}v_{j},
\end{equation}
where
\begin{equation}
	\label{5}
	\overline p=p-3H\xi
\end{equation}
 In above Eqs. (\ref{4}) and (\ref{5}) the different quantities have there usual meaning as already described in \cite{ref39}. 
 The four velocity of the particles $v^{i}=(0,0,0,1)$ and a unit space-like vector
$ x^{i} $  representing the direction of string satisfy  $g_{ij}v^{i}v^{j}=-g_{ij}x^{i}x^{j}=-1, v^{i} x_{i} =0$. In 
LRS Bianchi type-II metric, the mean Hubble parameter $H$ can be defined as

\begin{equation}
	\label{6}
	H=\frac{\dot a}{a}=\frac{1}{3}\left(2\frac{\dot X}{X}+\frac{\dot Y}{Y}\right)=\frac{1}{3}(2H_{1}+H_{2}).
\end{equation}
h
Here $H_{1}=\frac{\dot X}{X} $ and $ H_{2}=\frac{\dot Y}{Y}$ are directional Hubble parameters in the directions of $x$ and $y$ 
axes respectively. Here  $a=a(t)$ is average scale factor, which, for LRS B-type-II model, is written as

\begin{equation}
	\label{7}
	a(t)=(X^{2}Y)^{\frac{1}{3}}
\end{equation}
The  particle density denoted by $\rho_{\rho}$ follows the relation

\begin{equation}
	\label{8}
	\rho=\rho_{\rho}+\lambda
\end{equation}

For the Metric, (\ref{1}), the S$\acute{a}$ez-Ballester field equations (\ref{2}) \& (\ref{3}), along with energy-momentum tensor 
given by (\ref{4}), we obtain the following system of field equations

\begin{equation}
	\label{9}
	\frac{\ddot X}{X}+\frac{\ddot Y}{Y}+\frac{\dot X}{X}\frac{\dot Y}{Y}+\frac{1}{4}\frac{Y^{2}}{X^{4}}-
	\frac{1}{2}\omega\phi^{r}\dot \phi^{2}=\lambda-\overline p
\end{equation}

\begin{equation}
	\label{10}
	2\frac{\ddot X}{X}+\frac{\dot X^{2}}{X^2}-\frac{3}{4}\frac{Y^{2}}{X^{4}}-\frac{1}{2}\omega\phi^{r}\dot \phi^{2}=-\overline p
\end{equation}

\begin{equation}
	\label{11}
	\frac{\dot X^{2}}{X^2}+2\frac{\dot X}{X}\frac{\dot Y}{Y}-\frac{1}{4}\frac{Y^{2}}{X^{4}}-\frac{1}{2}\omega\phi^{r}\dot \phi^{2}=\rho
\end{equation}

\begin{equation}
	\label{12}
	\ddot \phi+\dot \phi\left(\frac{2\dot X}{X}+\frac{\dot Y}{Y}\right)+\frac{r}{2}\frac{\dot \phi^{2}}{\phi}=0
\end{equation}
In the usual notation, expansion scalar $\theta$ and the shear scalar ($\sigma$) are defined and given as  
\begin{equation}
	\label{13}
	\theta=v^{i}_{;j}=3\frac{\dot a}{a}=2\frac{\dot X}{X}+\frac{\dot Y}{Y}
\end{equation}
and
\begin{equation}
	\label{14}
	\sigma^{2}=\frac{1}{2}\sigma_{ij}\sigma^{ij}=\frac{1}{2}\left[2\frac{\dot X^2}{X^2}+\frac{\dot Y^2}{Y^2}\right]-\frac{1}{6}\theta^{2}
\end{equation}
where

$
\sigma_{ij}=v_{i;j}+\frac{1}{2}(v_{i;k}v^{k}v_{j}+v_{j;k}v^{k}v_{i})+\frac{1}{3}\theta(g_{ij}+v_{i}v_{j})$\\

The anisotropy parameter $(A_{m})$ is defined as

\begin{equation}
	\label{15}
	A_{m} = 6\left(\frac{\sigma}{\theta} \right)^{2} = \frac{2\sigma^{2}}{3H^{2}}
\end{equation}

\subsection{Assumptions}   

There are four equations (\ref{9})-(\ref{12}) having seven unknowns $X, Y, \phi, p, \rho, \xi $ and $ \lambda$. For
deterministic solutions of this system, we have to take three more equations, which relates these parameters.\\

As suggested by Thorne \cite{ref40} and followed by many researchers \cite{ref41,ref42}, we 
first, assume $\theta$ is proportional to $\sigma$ which gives

\begin{equation}
	\label{16}
	\frac{1}{\sqrt{3}}\left(\frac{2\dot X}{X}-\frac{\dot Y}{Y}\right)=\ell \left(\frac{2\dot X}{X}+\frac{\dot Y}{Y}\right)
\end{equation}
where $\ell$ is the constant of proportionality. This yields

\begin{equation}
\label{17}
\frac{\dot X}{X}=m \frac{\dot Y}{Y},
\end{equation}
where $ m=\frac{\sqrt{3}+\ell}{\ell -2\sqrt{3}}$ . We have select $ m > 0$ for anisotropic universe, provided $m \neq 1$, 
as the study presents a picture of FRW model for $m=1$. Integrating Eq. (\ref{17}) and we get
\begin{equation}
\label{18}
X=c_{1}(Y)^{m},
\end{equation}
where $c_{1}$ is a constant of integration. Any loss of generality and for simplicity, 
$c_{1}=1$ is considered. Hence Eq. (\ref{18}) is reduced to  
\begin{equation}
\label{19}
X=(Y)^{m}
\end{equation}

Secondly, we consider $q$ as linear function of Hubble parameter \cite{ref43}$-$\cite{ref46}:

\begin{equation}
\label{20} q = -\frac{a \ddot{a}}{\dot{a}^{2}} = \beta H + \alpha = \beta \frac{\dot{a}}{a} + \alpha.
\end{equation}
Here $\alpha$, and $\beta$ stand for arbitrary constants. Eq. (\ref{20}) renders as $ \frac{a \ddot{a}}{\dot{a}^{2}} + 
\beta \frac{\dot{a}}{a} + \alpha = 0$, which by solving proceeds as  

\begin{equation}
\label{21} a = exp{\left[-\frac{(1 + \alpha)}{\beta}t -\frac{1}{(1 + \alpha)} + \frac{l}{\beta}\right]}, ~ ~ provided ~ \alpha \ne -1.
\end{equation}
Here $l$ is a constant of integration. \\

From Eq. (\ref{21}), we calculate
\[
 \dot{a} = -\left(\frac{1 + \alpha}{\beta}\right)\exp{\left[-\left(\frac{1 + \alpha}{\beta}\right)t\right]} - 
 \frac{1}{(1 + \alpha)} + \frac{l}{\beta},
\]

\begin{equation}
\label{22} \ddot{a} = \left(\frac{1 + \alpha}{\beta}\right)^{2}\exp{\left[-\left(\frac{1 + \alpha}{\beta}\right)t - 
\frac{1}{(1 + \alpha)} + \frac{l}{\beta}\right]}.
\end{equation}

Eqs. (\ref{20}) and (\ref{22}) render the value of DP as $q = -1$. We also observed the same value of DP for 
$\alpha = 0$. \\

For $\alpha = -1$, Eq. (\ref{20}) is changed into the form:
\begin{equation}
\label{23} q = -\frac{a \ddot{a}}{\dot{a}^{2}} = -1 + \beta H ,
\end{equation}

Eq. (\ref{23}) reproduced the following differential equation:

\begin{equation}
\label{24} \frac{a\ddot{a}}{\dot{a}^{2}} + \beta \frac{\dot{a}}{a} -1 = 0.
\end{equation}

The solution of above equation is found to be (Sharma et al. 2019, Garg et al. 2019)  
\begin{equation}
\label{25} a = \exp{\left[\frac{1}{\beta}\sqrt{2\beta t + k}\right]},
\end{equation}
where $k$ is an integrating constant. Eq. (\ref{25}) is recently used by different authors 
\cite{ref43}$-$\cite{ref46} in different contexts.  \\

For the study of cosmic decelerated-accelerated expansion of the universe, we only consider the case 
$\alpha = -1$.\\

For the scale factor (\ref{25}), the DP $(q)$ and Hubble parameter $H$ is given as
\begin{equation}
\label{26}  q=-1+\frac{\beta}{\sqrt{2\beta t+k}}, ~ ~ H=\frac{1}{\sqrt{2\beta t+k}}.
\end{equation}
From Eq. (\ref{26}), we observe that $q>0$ for $t<\frac{\beta^{2}-k}{2\beta}$ and $q<0$ for 
$t>\frac{\beta^{2}-k}{2\beta}$. 
 

\section{Solution of the field Equation}
By using Eqs. (\ref{19}), (\ref{25}) and (\ref{7}), we obtain:
\begin{equation}
	\label{27}
	X=(e^{\frac{1}{\beta}\sqrt{2 \beta t + k}})^{\frac{3m}{2m+1}},
\end{equation}
\begin{equation}
	\label{28}
	Y=(e^{\frac{1}{\beta}\sqrt{2 \beta t + k}})^{\frac{3}{2m+1}}.	
\end{equation}
From Eqs. (\ref{12}), (\ref{27}) and (\ref{28}), we evaluate scalar field $(\phi)$ as
\begin{equation}
	\label{29}
	\phi(t)=\left[\frac{r+2}{2}\left(\phi_{0}\int {\frac{dt}{(e^{\frac{3}{\beta}\sqrt{2 \beta t +k}})}}+
	\phi_{1}\right)\right]^{\frac{2}{r+2}},
\end{equation}

where $\phi_{0}$ and $\phi_{1}$ are integrating constants.\\

Solving Eqs. (\ref{9})-(\ref{11}) by using Eqs. (\ref{27})-(\ref{29}), we obtain energy density $\rho $, effective 
pressure $\overline p$ and string tension density $\lambda $ as 
\begin{equation}
	\label{30}
	\rho=\Biggl[\frac{9m(m+2)}{(2m+1)^{2}(2 \beta t + k)} -\frac{1}{4}(e^{\frac{1}{\beta}\sqrt{2 \beta t + k}})^{\frac{6-12m}{2m+1}}
	+\frac{1}{2}\frac{\omega \phi_ {0}^{2}}{(e^{\frac{6}{\beta}\sqrt{2 \beta t + k}})}\Biggr],	
\end{equation}

\[
\overline p =\Biggl[-\frac{27m^{2}}{(2m+1)^{2}(2 \beta t + k)}+\frac{6m \beta}{(2m+1)}{(2\beta t+k)^{\frac{-3}{2}}}+
\]
\begin{equation}
	\label{31}
	\frac{3}{4}(e^{\frac{1}{\beta}\sqrt{2 \beta t + k}})^{\frac{6-12m}{2m+1}} +\frac{1}{2}\frac{\omega \phi_ {0}^{2}}
	{(e^{\frac{6}{\beta}\sqrt{2 \beta t + k}})}\Biggr],
\end{equation}

\[\lambda =\Biggl[\frac{-18m^{2} +9m+9}{(2m+1)^{2}(2 \beta t + k)}-\frac{3\beta(m-1)}{(2m+1)(2\beta t+k)^{\frac{3}{2}}}+
\]
\begin{equation}
	\label{32}
	(e^{\frac{1}{\beta}\sqrt{2 \beta t + k}})^{\frac{6-12m}{2m+1}}\Biggr].
\end{equation}
Accordingly, the particle density $\rho_{p}$ is obtained as

\[\rho_{p}=\Biggl[\frac{(27^{2}+9m-9)}{(2m+1)^{2}(2 \beta t + k)}-\frac{5}{4}(e^{\frac{1}{\beta}
\sqrt{2 \beta t + k}})^{\frac{6-12m}{2m+1}} +
\]
\begin{equation}
\label{33}
\frac{1}{2}\frac{\omega \phi_ {0}^{2}}{(e^{\frac{6}{\beta}\sqrt{2 \beta t + k}})}-\frac{3\beta (m-1)}{(2m+1){2\beta t+k)}^{\frac{3}{2}}}\Biggr].	
\end{equation}

For calculating the other parameters, we shall consider the following three cases.

\subsection{Case I: Bulk Viscous Model with $p=\alpha \rho$}

Considering perfect gas equation of state as:
\begin{equation}
	\label{34}
	p=\alpha{\rho},
\end{equation}
where $\alpha$ $(0 \leq \alpha \leq 1)$ is a constant. For the various values of $\alpha$, we will get three types of models:\\

(i) if $\alpha=0$, we tend to get matter dominant model.\\

(ii) if $\alpha=\frac{1}{3}$, we tend to get radiation dominant model. \\

if $\alpha=1$, we get $\rho$=$p$ which is termed as Zel'dovich fluid or stiff fluid model \cite{ref47}. \\

Therefore by Eqs. (\ref{5}) and (\ref{34}), we can directly calculate the following values of $(p)$ and  $(\xi)$ :

\begin{equation}
\label{35}
p=\Biggl[\frac{9m\alpha(m+2)}{(2m+1)^{2}(2 \beta t + k)} -\frac{\alpha}{4}(e^{\frac{1}{\beta}\sqrt{2 \beta t + k}})^{\frac{6-12m}{2m+1}}
+\frac{\alpha}{2}\frac{\omega \phi_ {0}^{2}}{(e^{\frac{6}{\beta}\sqrt{2 \beta t + k}})}\Biggr]	
\end{equation}

\[\xi=\Biggl[\frac{(3m^{2}\alpha+6m\alpha+9m^{2})}{(2m+1)^2\sqrt{2\beta t+k}}-(\frac{\alpha+3}{12})(e^{\frac{1}
{\beta}\sqrt{2 \beta t + k}})^{\frac{6-12m}{2m+1}}\sqrt{2 \beta t + k}-
\]
\begin{equation}
\label{36}
\frac{2m\beta}{(2m+1)(2\beta t+k)}+(\frac{\alpha-1}{6})\frac{\omega\phi_{0}^{2}\sqrt{2\beta t+k}}{(e^{\frac{6}{\beta}\sqrt{2\beta t+k})}}\Biggr]
\end{equation}

\subsection{Case II: Bulk Viscous Model with $\xi=\xi_{0}\rho^{n}$}

For most of the investigations, we found that the coefficient of bulk 
viscosity $\xi$ is considered as a simple power function of energy density and it depends on time. It is assumed, 

\begin{equation}
	\label{37}
	\xi=\xi_{0}\rho^{n},
\end{equation}

where $\xi_{0}$ and $n$ are real constants \cite{ref48}$-$\cite{ref50}. For small density 
and radiative fluid, $n$ may be equal to $1$ \cite{ref51,ref52}.  For $(0 \leq n \leq 1/2)$ is good assumption to obtain 
realistic results, as given by Belinskii and Khalatnikov (1975). \\ 

 Using Eqs. (\ref{5}), (\ref{30}), (\ref{31}) and (\ref{37}), the expressions for  $\xi$ and  $p$ are given as:
\begin{equation}
\label{38}
\xi=\xi_{0}\left[\frac{9m(m+2)}{(2m+1)^{2}(2 \beta t + k)} -\frac{1}{4}(e^{\frac{1}{\beta}\sqrt{2 \beta t + k}})^{\frac{6-12m}{2m+1}} 
+\frac{1}{2}\frac{\omega \phi_ {0}^{2}}{(e^{\frac{6}{\beta}\sqrt{2 \beta t + k}})}\right]^{n}
\end{equation}

\[p= \Biggl[\frac{3\xi_{0}}{\sqrt{2\beta t+k}}\Biggl[\frac{9m(m+2)}{(2m+1)^{2}(2 \beta t + k)} -\frac{1}{4}(e^{\frac{1}{\beta}
\sqrt{2 \beta t + k}})^{\frac{6-12m}{2m+1}} + 
\]
\[\frac{1}{2}\frac{\omega \phi_ {0}^{2}}{(e^{\frac{6}{\beta}\sqrt{2 \beta t + k}})}\Biggr]^{n}-\frac{27m^{2}}{(2m+1)^{2}
(2 \beta t + k)}+ \frac{6m \beta}{(2m+1)}{(2\beta t+k)^{\frac{-3}{2}}}	
\]
\begin{equation}
	\label{39}
	+\frac{3}{4}(e^{\frac{1}{\beta}\sqrt{2 \beta t + k}})^{\frac{6-12m}{2m+1}} +\frac{1}{2}
	\frac{\omega \phi_ {0}^{2}}{(e^{\frac{6}{\beta}\sqrt{2 \beta t + k}})}\Biggr]
\end{equation}

\subsection{Case III: Perfect Fluid Model with $\xi=0$}

For perfect fluid, the coefficient of bulk viscosity is assumed to be zero. The 
rest of six unknowns $ X, Y, \phi, p, \rho$ and $\lambda$ can be directly calculated from the field Eqs. (\ref{9})-(\ref{12}). For
$ \xi=0$, Eq. (\ref{5}) gives $\overline{p}=p$ i.e. effective pressure equals to isotropic pressure, and the expression is given by

\[p =\Biggl[-\frac{27m^{2}}{(2m+1)^{2}(2 \beta t + k)}+\frac{6m \beta}{(2m+1)}{(2\beta t+k)^{\frac{-3}{2}}}	+
\]
\begin{equation}
\label{40}
\frac{3}{4}(e^{\frac{1}{\beta}\sqrt{2 \beta t + k}})^{\frac{6-12m}{2m+1}} +\frac{1}{2}
\frac{\omega \phi_ {0}}{(e^{\frac{6}{\beta}\sqrt{2 \beta t + k}})}\Biggr]
\end{equation}

\section{Interpretation of the Results}

From Eq. (\ref{26}), the present value of declaration parameter can be taken as $q_{0}=-1+\beta H_{0} = -1+
\frac{\beta}{\sqrt{2\beta t_{0}+k}}$, where $ H_{0}$ and $ t_{0}$ have their usual meaning.\\

By using the recent constraints ($H_{0}=73.8$, and $q_{0} = -0.54$) from SN Ia data in combination BAO and CMB observations 
\cite{ref54}, we concentrate the values of $\beta = 0.0062$ and $k = 0.000016$. We have used these values in 
formulating and drawing the different figures to analyze the nature of physical quantities. \\


\begin{figure}[H]
	(a)\includegraphics[width=7cm,height=5cm,angle=0]{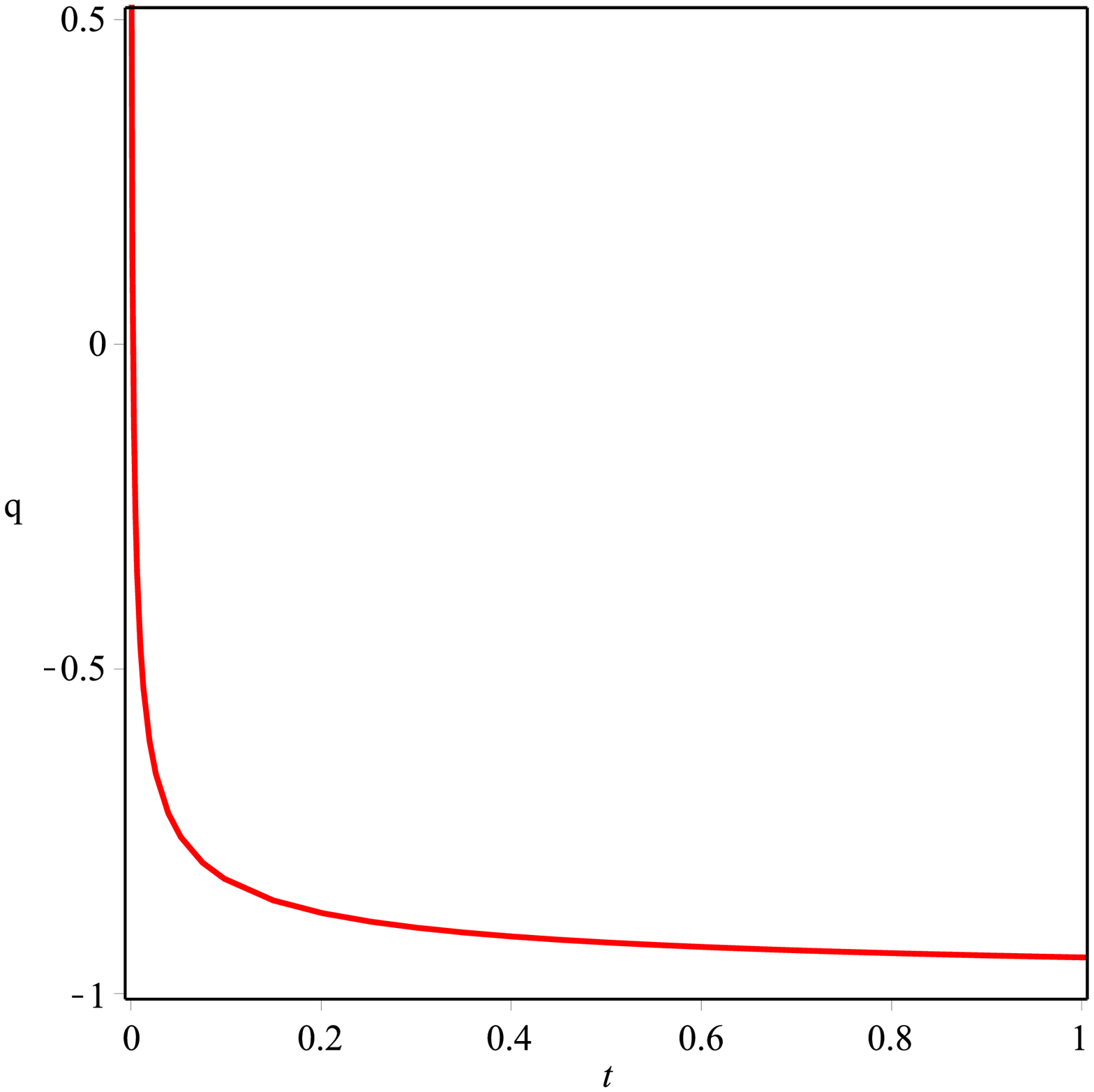}
	(b)\includegraphics[width=7cm,height=5cm,angle=0]{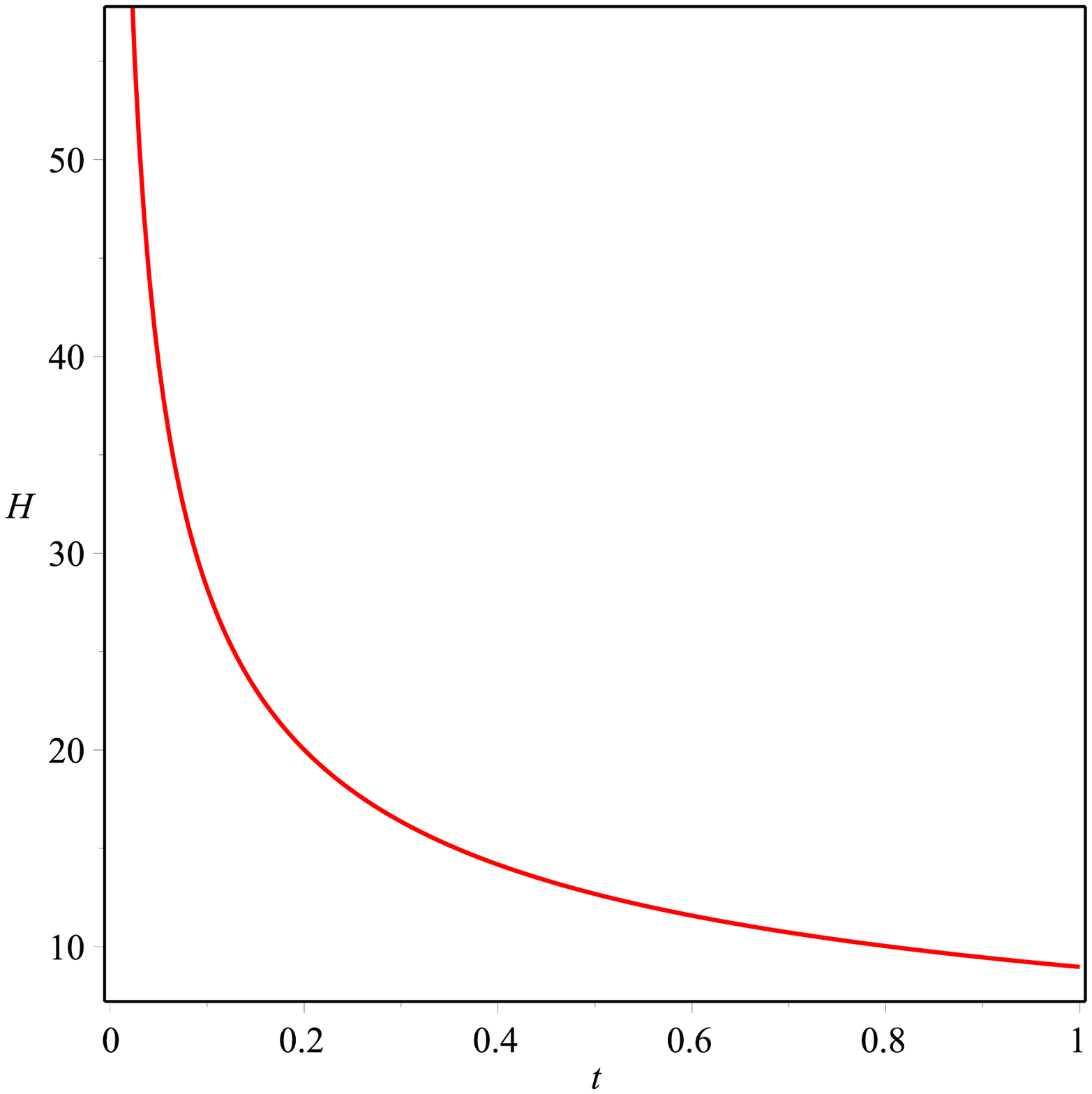}
	\caption{(a) The plot DP $q$ versus  $t$, (b) The plot of 
	$H$ versus $t$, Here $\beta=0.0062$ and $k=0.000016$.}.
\end{figure}

We have plotted the variation of deceleration parameter $q$ with respect to cosmic time $t$ in Figure $1(a)$, and observed 
that deceleration parameter is positive at early time and negative at present time indicating that our models are evolving 
from decelerating phase ($q>0$) to accelerating phase ($q<0$), and the models show a phase transition from positive to negative 
for DP $q$ for $k$ = $0.000016$ and $\beta$ = $0.0062$. The critical time at which the phase transaction took place is given 
by $t_c=\frac{\beta^{2}-k}{2\beta}$. Also, when $t \to \infty,~ q \to -1$. According to SNe Ia observation, 
the universe is accelerating at present and the value of DP lies in the range $-1 < q < 0$. So our models 
show consistency with recent observations.\\

Figure $1(b)$ shows the variation of Hubble parameter $H$ with respect to cosmic time $t$ as per Eq. (\ref{26}). 
We see that $H$ is a positive, decreasing function of time, and tends to zero as $t \to \infty$, which totally 
agrees with the established theories.\\

The average scale factor $a(t)$ in terms of redshift $z$ is given by $a(t)=\frac{a_0}{1+z}$,
where $a_0$ is the present value of the average scale factor $a(t)$.\\

From Eq. (\ref{25}), we can get 
$a_0 = \exp{\left[\frac{1}{\beta}\sqrt{2\frac{\beta}{H_0} + k}\right]}$. Using the values of $\beta=0.0062, k=0.000016$ 
and $H_{0}=73.8$ we get $a_{0}=8.6021$. We have used these values to draw the graph.\\

From Eq.(\ref{25}), we get $\frac{\sqrt{2\beta t + k}}{\beta}= ln(a)$. Also from $a(t)=\frac{a_0}{1+z}$, 
we have $ln(a)= ln(a_0)-ln(1+z)$.\\

Substituting the above in Eq. (\ref{21}), we get 
\begin{equation}
	\label{41}
	q(z)=-1+\frac{1}{ln(a_0)-ln(1+z)}
\end{equation} 
.
\begin{figure}[H]
	(a)\includegraphics[width=7cm,height=5cm,angle=0]{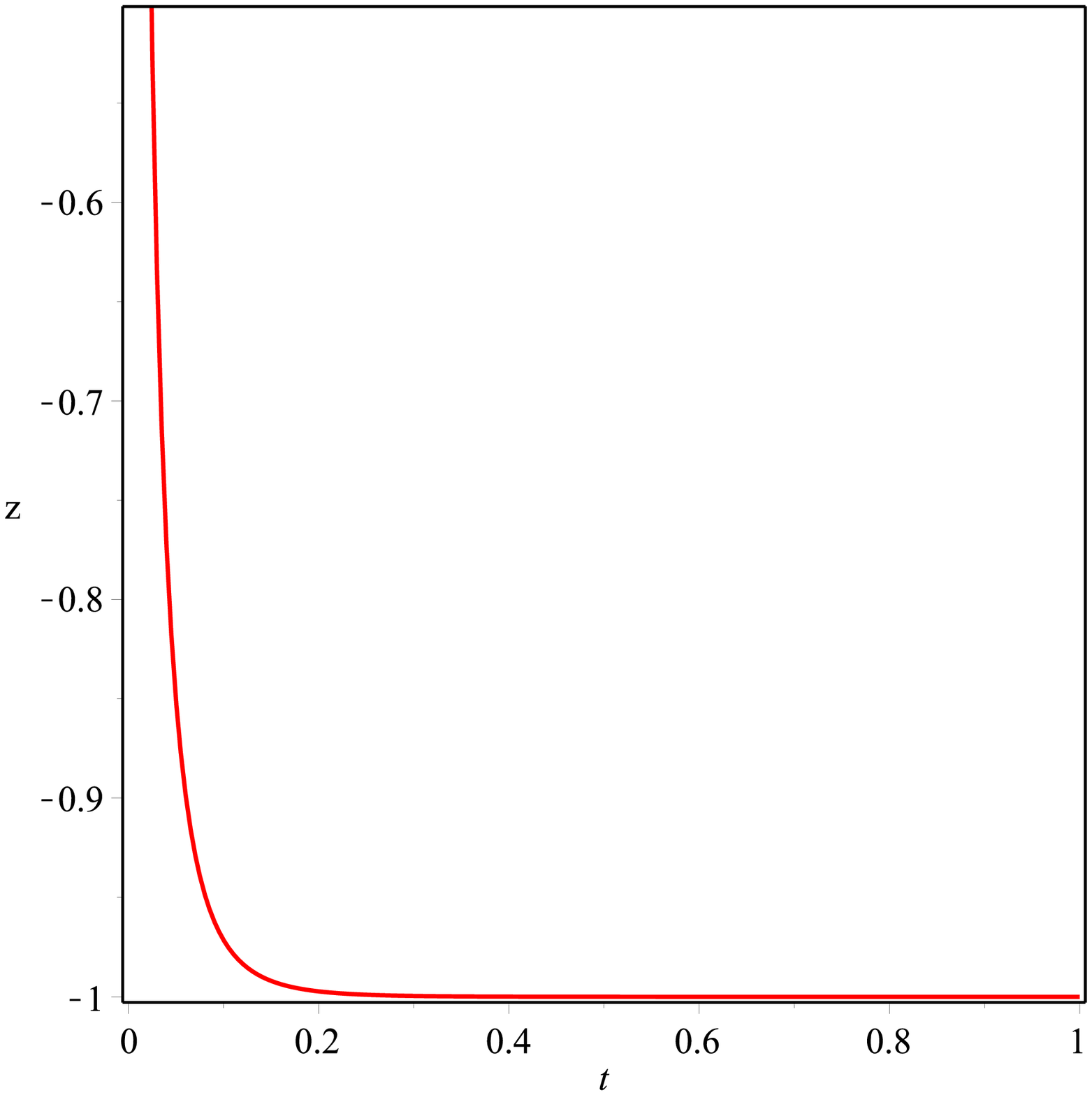} 
	(b)\includegraphics[width=7cm,height=5cm,angle=0]{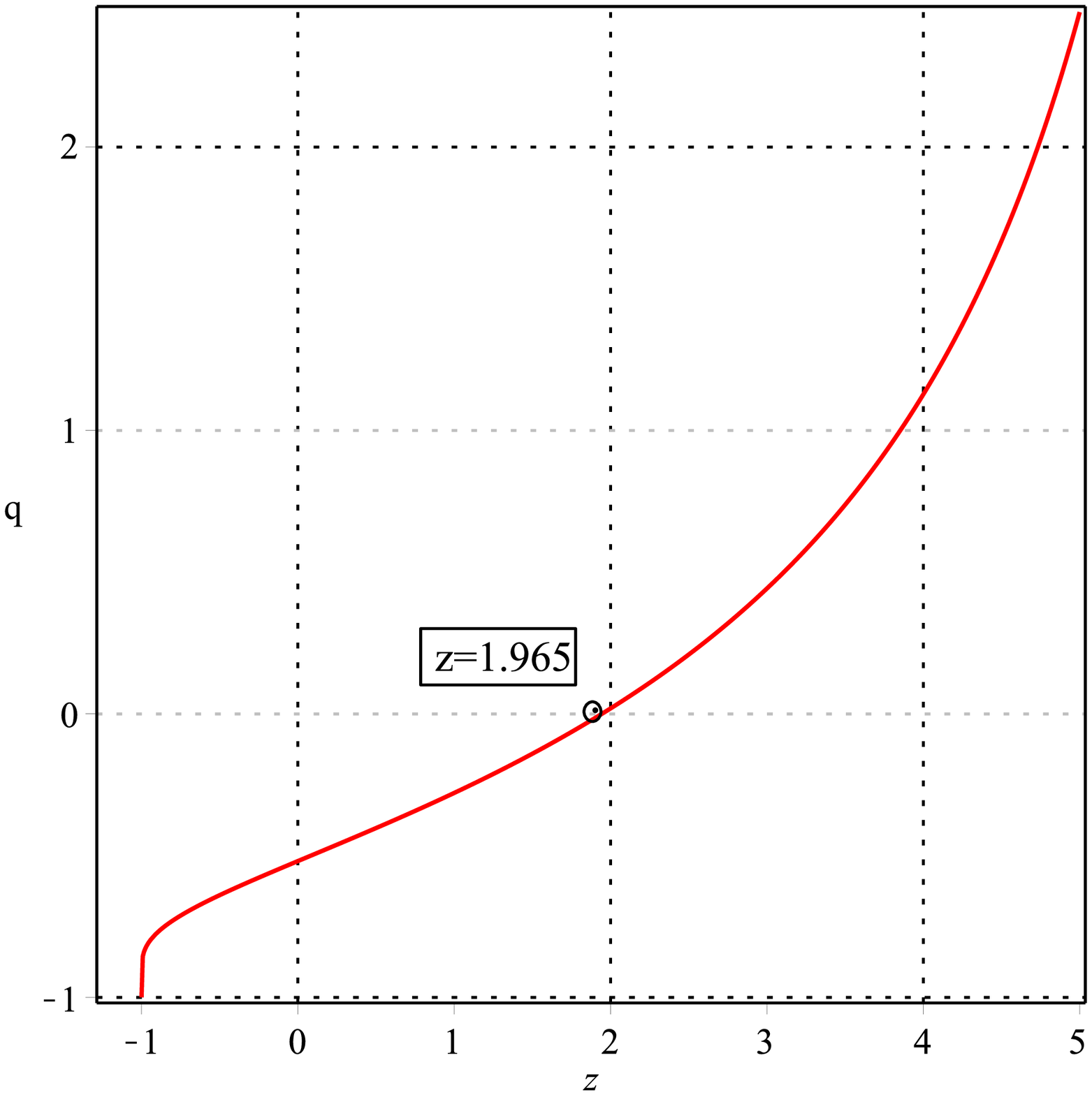}
	\caption{(a) The plot of redshift $z$ versus cosmic time $t$, (b) The plot of deceleration parameter $q$ versus redshift $z$,}.
\end{figure}
Figure $2(a)$ shows the fluctuation of redshift $z$ with cosmic time $t$ for our derived models. From the figure, we see that
the redshift $z$ is a monotonic decreasing function of cosmic time $t$ for the present value $a_{0}=8.6021$.
Also, $z$ starts with a small positive value $5.16$ at $t=0$ and $z \to -1$ as $t \to \infty$ for our derived models. So, we can
say that $t \to \infty$ corresponds to $z \to -1$\\  

In Figure $2(b)$, we have shown the fluctuation of $q$ concerning for redshift $z$ as per Eq. (\ref{41}). 
From this figure, we see that as the redshift $z$ decreases, the DP $q$ is changing its phase from positive 
(decelerating phase) to negative (accelerating phase) and $q \to -1$ as $z \to -1$.
Recently, \cite{ref55,ref56} have studied the transition redshift 
in $f(T)$ cosmology and observational constraints and cosmographic bounds on the cosmological deceleration-acceleration 
transition redshift in $F(R)$ gravity respectively. \\

It was found in its analysis that the SNe data favor current acceleration ($z < 0.5$) 
and past deceleration ($z > 0.5$). Recently, according to the High-z Supernova Search (HZSNS) team 
$z_{t} = 0.46 \pm 0.13$ at $(1\;\sigma)$ c.l. \cite{ref3} which has been further analyzed to 
$z_{t} = 0.43 \pm 0.07$ at $(1\;\sigma)$ c.l. \cite{ref3}. According to SNLS \cite{ref57}, 
as well as the one recently compiled by \cite{ref58}, yield a transition redshift $z_{t} \sim 0.6 (1\; \sigma)$ in 
better agreement with the flat $\Lambda$CDM model ($z_{t} = (2\Omega_{\Lambda}/\Omega_{m})^{\frac{1}{3}} - 1 \sim 0.66$). 
Another limit is $0.60 \leq z_{t} \leq 1.18$ ($2\sigma$, joint analysis) \cite{ref59}. Further, the transition 
redshift for our derived model comes to be $z_{t} \cong 1.965$ (see Fig. $2b$) which is in good agreement with the Type 
Ia supernovae observations, including the farthest known supernova SNI997ff at $z \approx 1.7$ \cite{ref2} and \cite{ref60}. 
We see that the variation of $q$ versus $z$ obtained in our model is compatible with the results obtained 
in the above references.\\

\begin{figure}[H]
	(a)\includegraphics[width=5.5cm,height=4cm,angle=0]{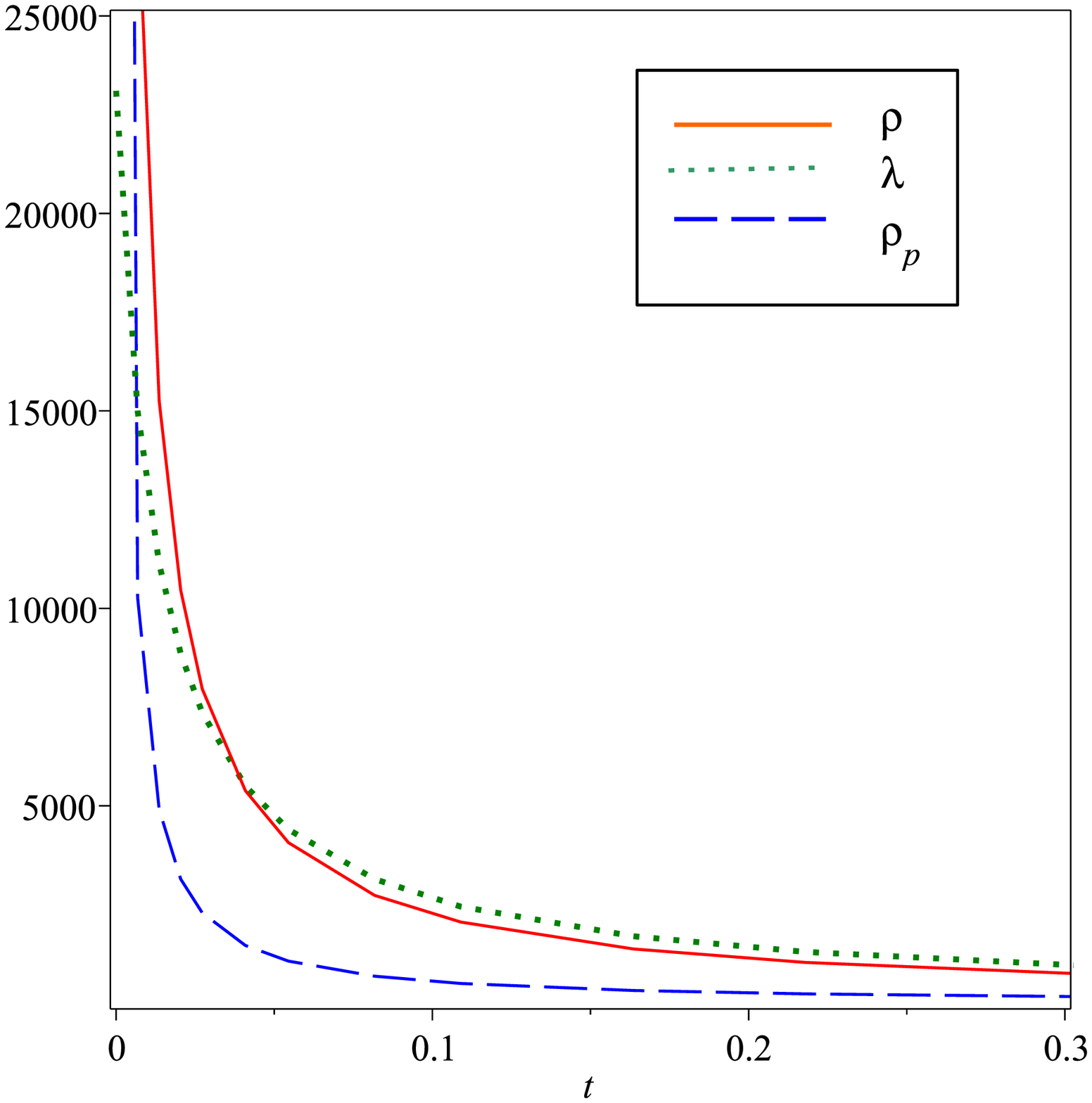}
    (b)\includegraphics[width=5.5cm,height=4cm,angle=0]{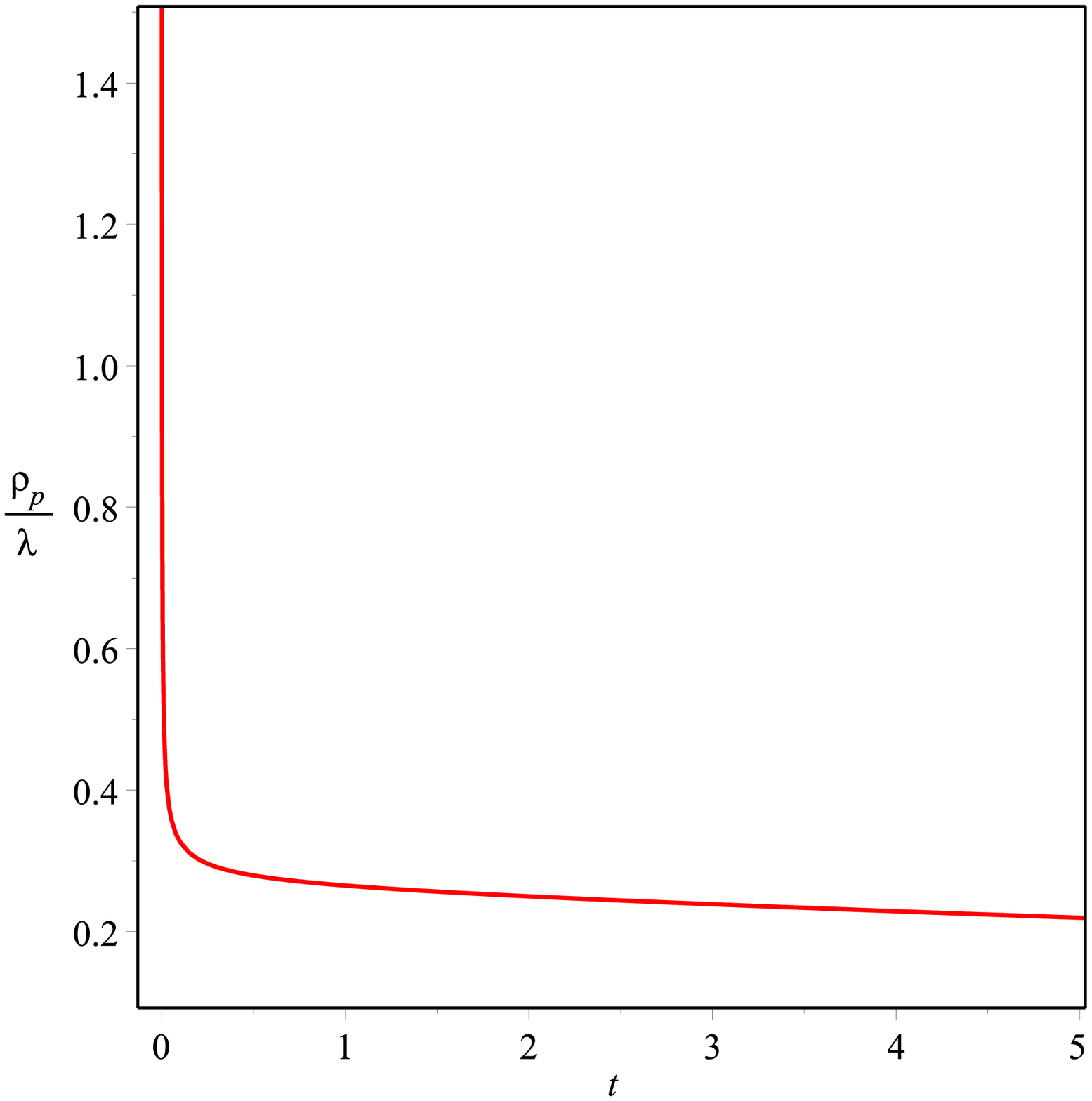}
    (c)\includegraphics[width=5.5cm,height=4cm,angle=0]{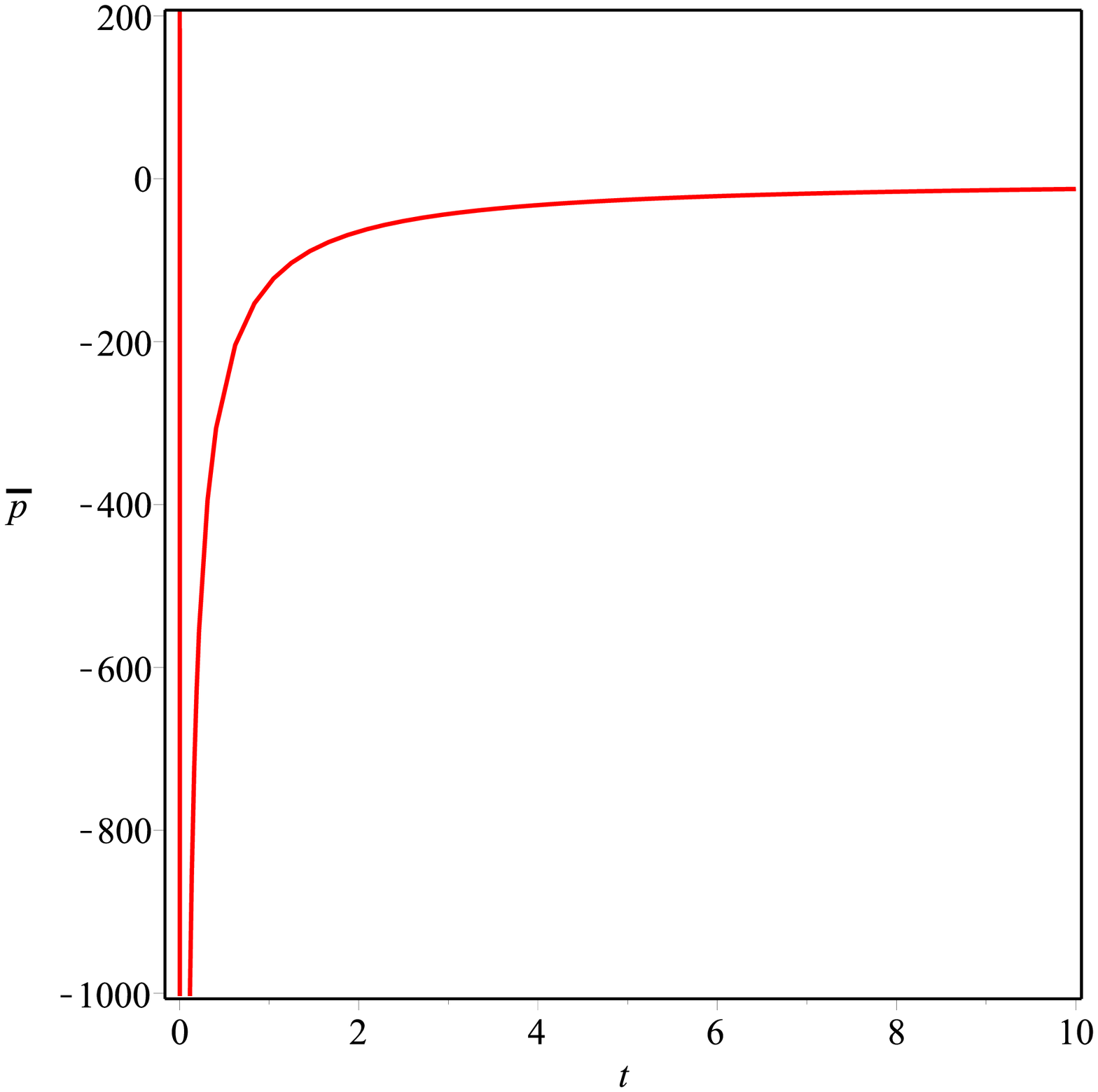}
	\caption{(a) Plot of energy densities versus cosmic time $t$, (b) The plot of $\frac{\rho_{p}}{\lambda}$ 
versus cosmic time $t$, (c) The plot of effective pressure $\overline{p}$ versus cosmic time $t$. Here $\omega=\phi_{0}=1, m=0.5, 
\beta=0.0062$ and $k=0.000016$}.
\end{figure}
In figure $3(a)$, we have shown three curves of total energy density $\rho$, particle energy density $\rho_p$ and string 
tension density $\lambda$. We see that all the energy densities are positive decreasing functions of time showing expanding 
universe. All the energy densities approach to zero as $t \to \infty$, meaning that the universe will keep on expanding 
forever. Also, we see that $\lambda < \rho_p$ for an early phase of the evolution i.e., particle dominates over the string, and then 
$\lambda > \rho_p$ in due course of evolution i.e., the string dominates over the particle  thereafter. \\

The comparative behavior of particle density $\rho_p$ and string tension density $\lambda$ is also studied in figure $3(b)$. 
From figure $3(b)$ we see that the ratio $\frac{\rho_{p}}{\lambda} > 0$ throughout. At the early phase of the evolution, 
the ratio $\frac{\rho_{p}}{\lambda} > 1$ , indicating that $\rho_{p} >\lambda$ i.e. the particle dominated phase. But, as the 
time progresses, the ratio falls below $1$ indicating the string dominated phase. These observations are supported by Krori \cite{ref61} 
and Kibble \cite{ref62}.\\

In Figure $3(c)$, we have plotted the variation of the effective pressure $\overline{p}$ concerning cosmic time $t$ as 
per Eq. (\ref{31}). We see that $\overline{p}$ is negative at present, which may be seed for current accelerated expansion 
of the universe.  \\

In Figure $4(a)$ we have plotted the behavior of isotropic pressure $p$ for case-I when $p=\alpha \rho$ for three scenarios 
$\alpha=0$ (dust filled), $\alpha=1/3$ (radiation dominated) and $\alpha=1$ (stiff matter filled) universe. In all the cases 
we find that the isotropic pressure is a positive decreasing function of time.\\

In Figure $4(b)$ we consider case-II when $\xi=\xi_{0}\rho^{n}$ and plotted $p(t)$ for three values of $n=0, 1/2$ and $1$. 
We observed that for all the three values of $n$, the isotropic pressure $p$ is again a positive decreasing function of time. 
We also observed in both the case-I and case-II that $p \to 0$ when $t \to \infty$.\\

In case-III, when $\xi=0$ (i.e., in the absence of viscous effect) the isotropic and effective pressures become equal. The behavior 
of the effective pressure is graphed in Figure $4(c)$. We see that in the absence of viscosity the effective pressure becomes highly 
negative at the early time then increases and tends to a small negative value at late time.

\begin{figure}[H]
	(a)\includegraphics[width=5.5cm,height=4cm,angle=0]{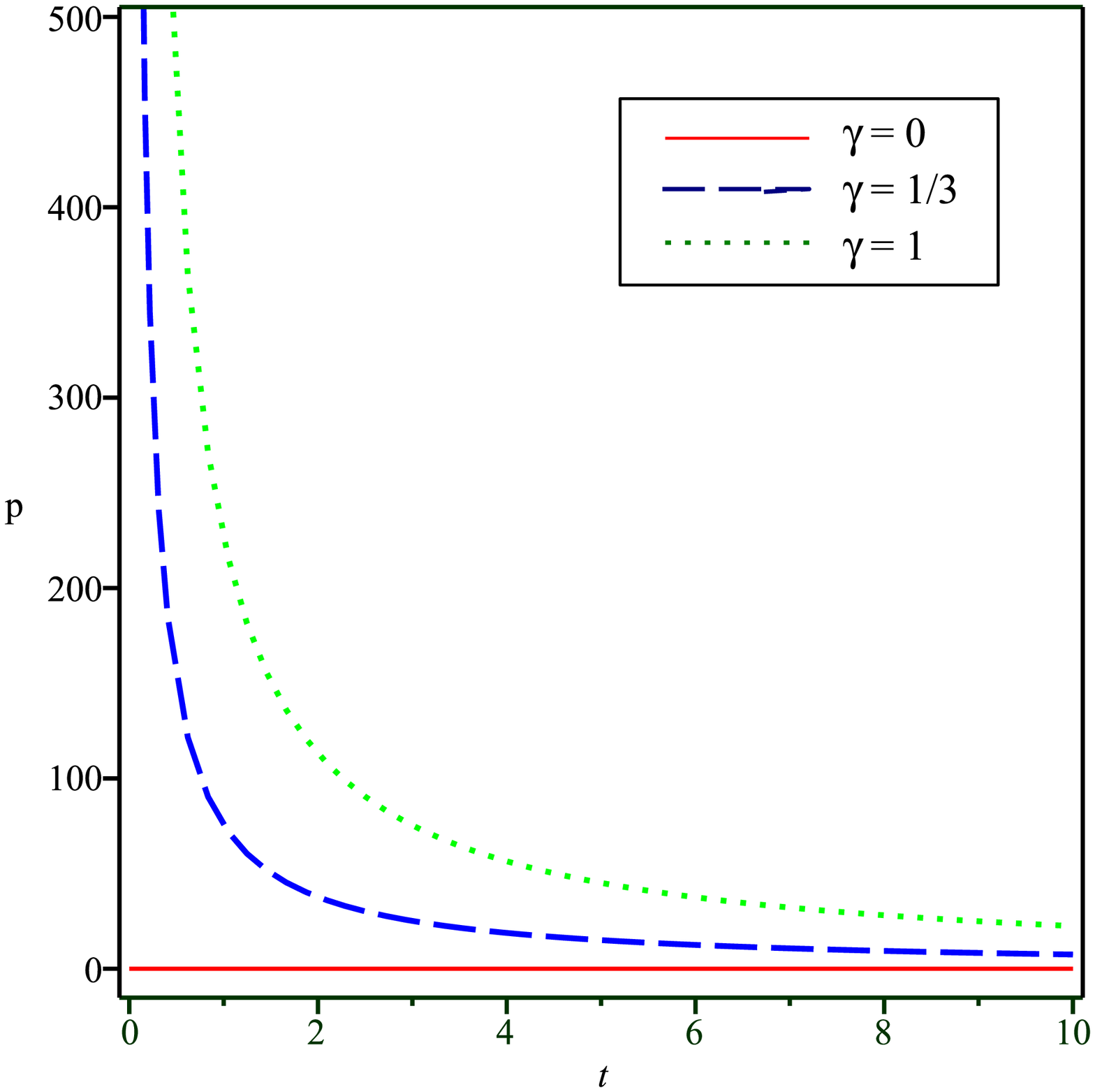}
	(b)\includegraphics[width=5.5cm,height=4cm,angle=0]{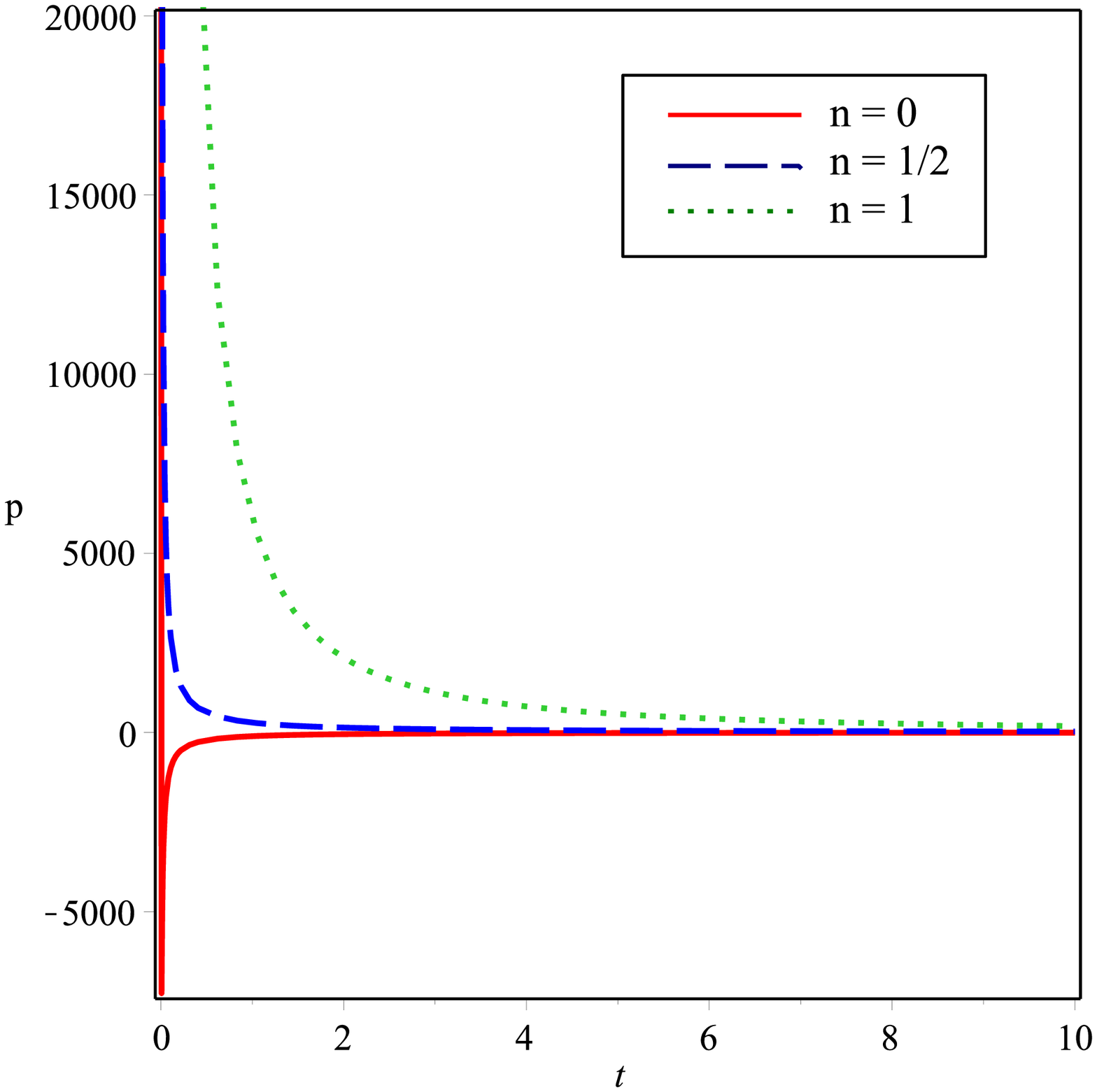}
	(c)\includegraphics[width=5.5cm,height=4cm,angle=0]{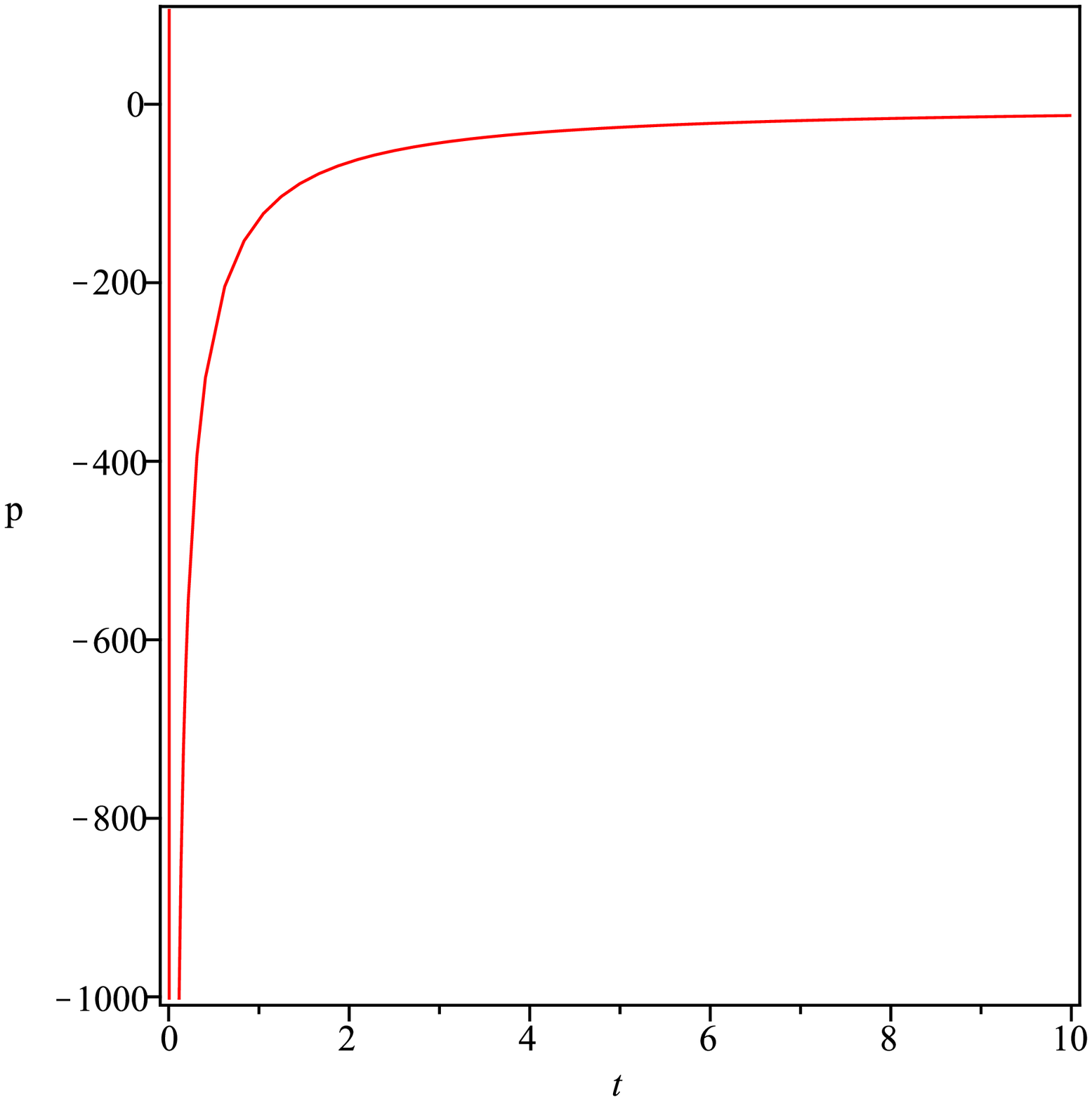}
	\caption{(a) Plot of isotropic pressure $p$ versus $t$ for case I, (b) Plot of isotropic pressure $p$ versus $t$ 
	for case II, (c) Plot of isotropic pressure $p$ versus $t$ for case III. Here $\omega=\phi_{0}=1, m=0.5, \beta=0.0062$ 
	and $k=0.000016$. }
\end{figure}
In Figure $5(a)$ we have plotted the of bulk viscosity coefficient $\xi$ for case-I when $p=\alpha \rho$ for three 
scenarios $\alpha=0, 1/3 ~\&~ 1$. In all the cases we find that $\xi$ is a positive decreasing function of time. In 
the early universe, it was high and after that it reduces gradually and tends to zero as $t \to \infty$. So, we can 
say that the nature of the fluid was highly viscous at the time of the early universe which tends to reduce and vanish in due 
course of time. In Figure $5(b)$ we consider case-II when $\xi=\xi_{0}\rho^{n}$ and plotted $\xi(t)$ for three values 
of $n=0, 1/2$ \& $1$. Here also, we observe the same behavior for the two values of $n=1/2$ and $1$, whereas for $n=0$ 
the viscous effect vanishes throughout the evolution of the universe. \\

\begin{figure}[H]
	(a)\includegraphics[width=7cm,height=5cm,angle=0]{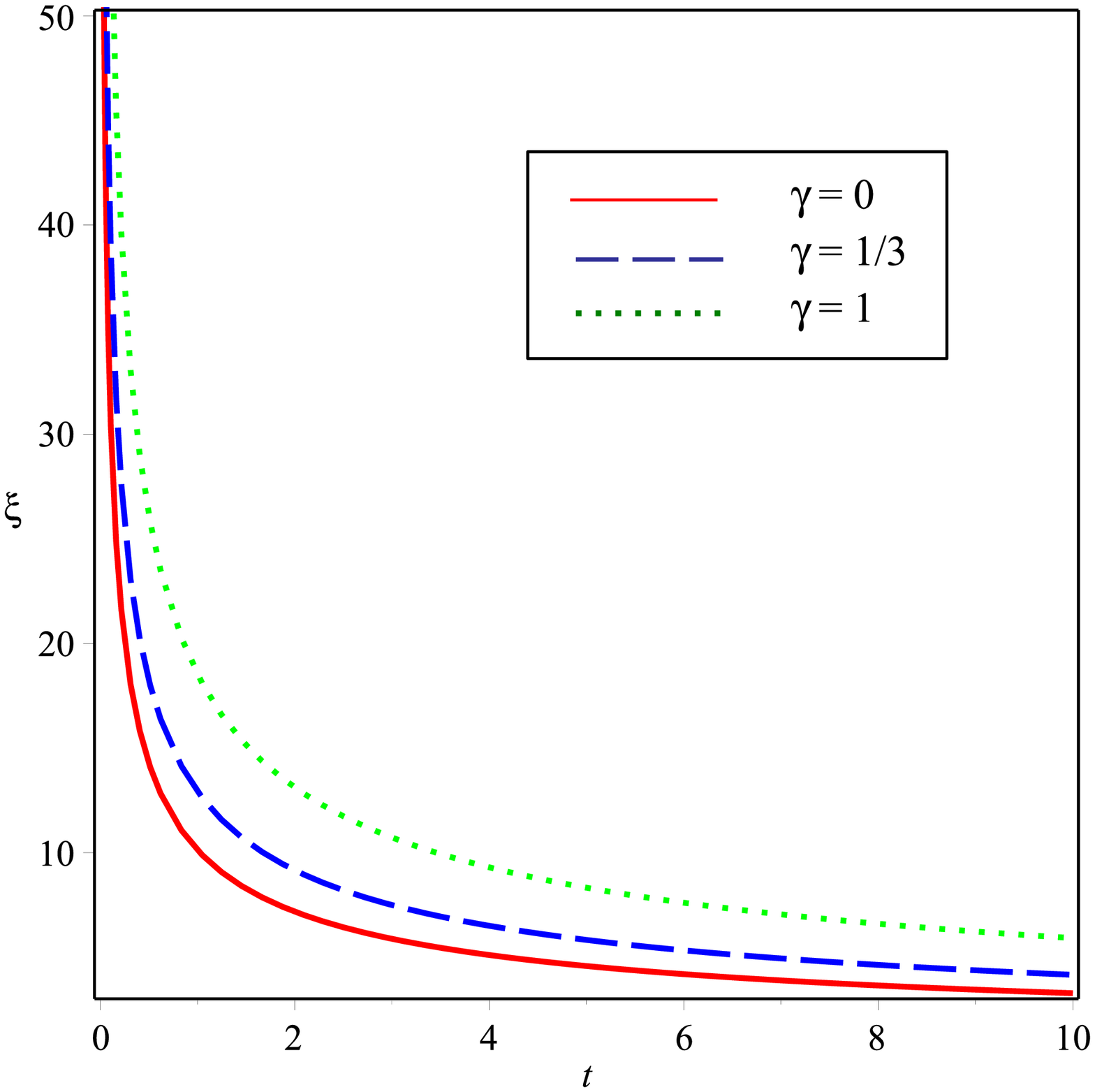}
	(b)\includegraphics[width=7cm,height=5cm,angle=0]{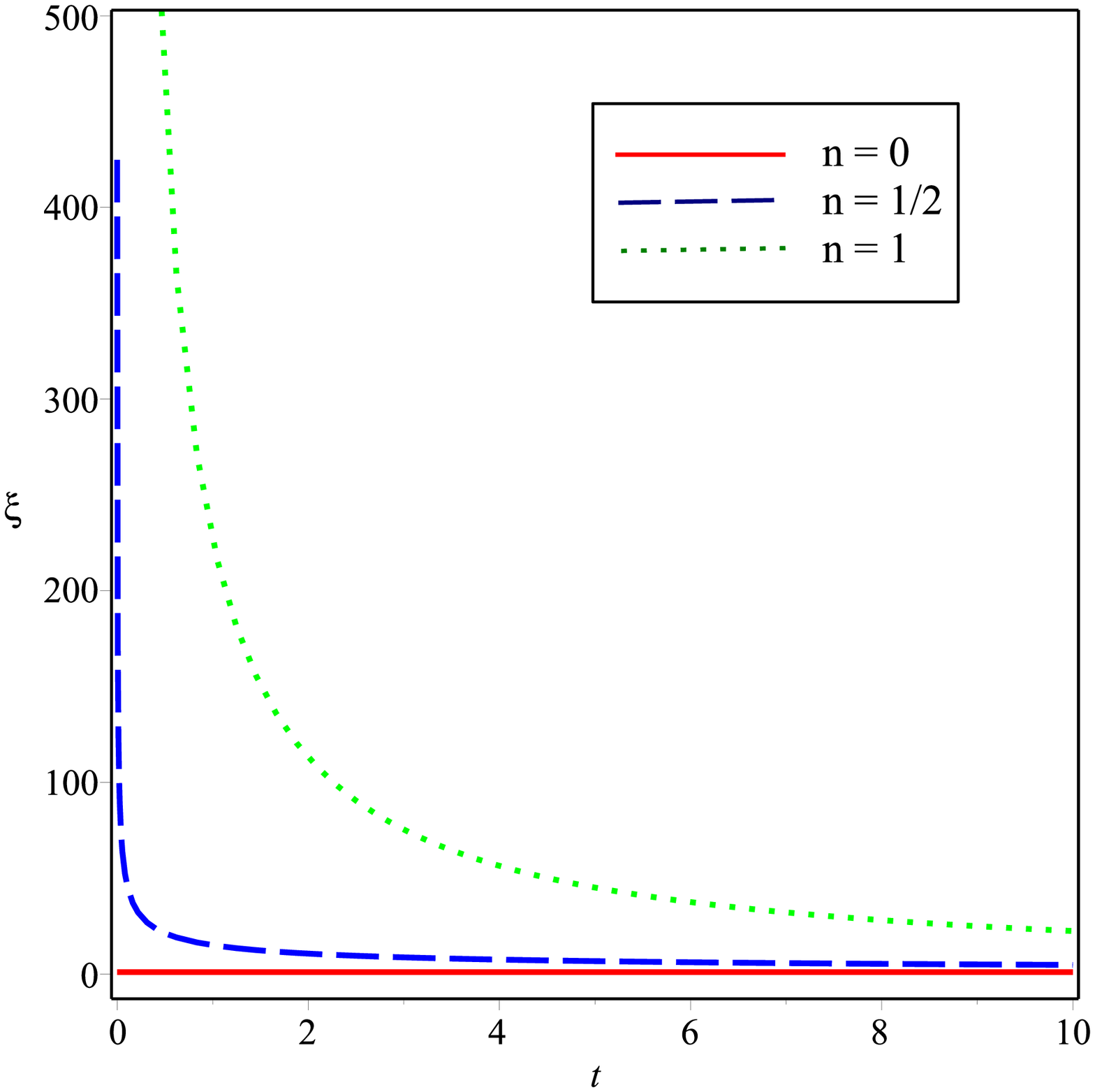}
	\caption{(a) Plot of viscosity parameter $\xi$ versus $t$ for case I, (b) Plot of viscosity parameter $\xi$ 
	versus $t$ for case II. Here $\omega=\phi_{0}=1, m=0.5, \beta=0.0062$ and $k=0.000016$}
\end{figure}
Other physical parameters expansion scalar($\theta)$, Volume scalar  $(V)$, shear scalar $\sigma$ and anisotropy 
parameter$(A_{m})$ and directional Hubble parameters $(H_{1}$ and $H_{2})$ are obtained as

\begin{equation}
	\label{42}
	\theta=3H=\frac{3}{\sqrt{2\beta t+k}}
\end{equation}
\begin{equation}
	\label{43}
	V=A^{2}B=e^{\frac{3}{\beta}\sqrt{2\beta t+k}}
\end{equation}
\begin{equation}
	\label{44}
	\sigma^{2}=-\frac{1}{2}\left[\frac{9(2m^{2}+1)}{(2m+1)^{2}(2\beta t+k)}\right]-\frac{3}{2\sqrt{2\beta t+k}}
\end{equation}

\begin{equation}
	\label{45}
	A_{m}=\frac{2m^{2}-4m+2}{(2m+1)^{2}}
\end{equation}

\begin{equation}
	\label{46}
	H_{1}=mH_{2}=\frac{3m}{(2m+1)\sqrt{2\beta t+k}}
\end{equation}

In Big Bang scenario all the  parameters like shear scalar ($\sigma$), expansion scalar ($\theta$), and Hubble parameter 
$(H)$ are finite. From Eq. (\ref{43}) Spatial volume $(V)$ is zero at $t=0$. As $t\to \infty$, $V$ becomes infinite whereas 
$\theta$, $H$, and $\sigma$ approach to zero.\\
\section{The Model Stability}
We have tested the stability of the background solution w.r.to perturbations of the metric. 
For the study, we adopt the notation $a_i$ for the metric potentials. (ie. $a_{1}=A$ and $a_{2}=B$).\\

The stability analysis is performed against the perturbations of all possible fields. The stability of the 
solution has been first discussed by Chen and Kao \cite{ref63}. Here perturbation will be considered for the two expansion factor $a_{i}$ via
\begin{equation}
	\label{47}
	a_{i}\rightarrow a_{B_{i}}+\delta {a_{i}}=a_{B_{i}}(1+\delta b_{i}). 
\end{equation}
where $\delta a_{i}=a_{B_{i}}{\delta b_{i}}$.\\

Accordingly, the perturbations of the volume scalar, directional Hubble factors, the mean Hubble parameter are shown as follows:
\[V \rightarrow V_{B} + V_{B} \sum_{i=1}^3 \delta b_{i},~~~
H_{i} \rightarrow H_{B_{i}} + \delta \dot b_{i},
\]
\begin{equation}
	\label{48}
	H\rightarrow H_{B}+\frac{1}{3}\sum_{i=1}^3 \delta \dot b_{i}, ~~~
	\sum_{i=1}^3 H_{i}^{2} \rightarrow \sum_{i=1}^3 H_{{B}_i}^{2}+2 \sum_{i=1}^3 H_{{B}_i}. \delta b_{i} 
\end{equation}
It can be derived that metric perturbations $\delta {b_{i}}$ is to linear order in  $\delta{b_{i}}$ obey the following equations
\begin{equation}
	\label{49}
	\Sigma \delta \ddot {b_{i}}+2\Sigma H_{B_{i}} \delta \dot b_{i}=0
\end{equation}

\begin{equation}
	\label{50}
	\Sigma \delta \ddot{b_{i}}+2\frac{\dot V_{B}}{V_{B}}{\delta \dot{b_{i}}}+\Sigma \delta \dot b_{j}H_{B_{i}}=0
\end{equation}

\begin{equation}
	\label{51}
	\Sigma \delta \dot b_{j}=0
\end{equation}

From the above three equations, It can easily be seen that
\begin{equation}
	\label{52}
	\delta \ddot{b_{i}}+\frac{\dot V_{B}}{V_{B}}\delta \dot b_{i}=0
\end{equation}
where $ V_{B}$ is the background volume scalar and in this model, it is given by

\begin{equation}
	\label{53}
	V_{B}=e^{\frac{3}{\beta}\sqrt{2\beta t+k}}
\end{equation}

We can calculate the $\delta{b_{i}}$ with the help of Eq. (\ref{51}), then we get

\begin{equation}
	\label{54}
	\delta b_{i} = e^{\frac{-3}{\beta}\sqrt{2\beta t+k}}\left(-{\frac{1}{3}\sqrt{2\beta t+k}}-\frac{1}{9}\beta\right)+c
\end{equation}

here $c$ is constant of integration. Therefore the actual fluctuations for each expansion factor 
$\delta a_{i}=a_{B_{i}}\delta{b_{i}}$ are given by
\begin{equation}
	\label{55}
	\delta a_{i} = a_{B_{i}}e^{\frac{-3}{\beta}\sqrt{2\beta t+k}}\left(-{\frac{1}{3}\sqrt{2\beta t+k}}-\frac{1}{9}\beta\right)+c
\end{equation}
From Eq. (\ref{55}) and  Figure $(6)$, we observed that for positive value of $\beta$ and $k,~  \delta{a_{i}}$ approaches 
to zero for large $t$ i.e. $t\to \infty , \delta a_{i} \to 0$. Consequently, the background solution is stable against the 
metric perturbation.\\

If $\delta{b_{i}}$ tends to zero, from Eq. (\ref{48}), we see that $V \to V_{B}, ~ H_{i} \to H_{B_i}, ~H \to H_{B}$ so we 
can say that our solution is stable against the perturbation of volume scale, directional Hubble and average Hubble parameters also. 

\begin{figure}[H]
	\centering
	\includegraphics[width=8cm,height=6cm,angle=0]{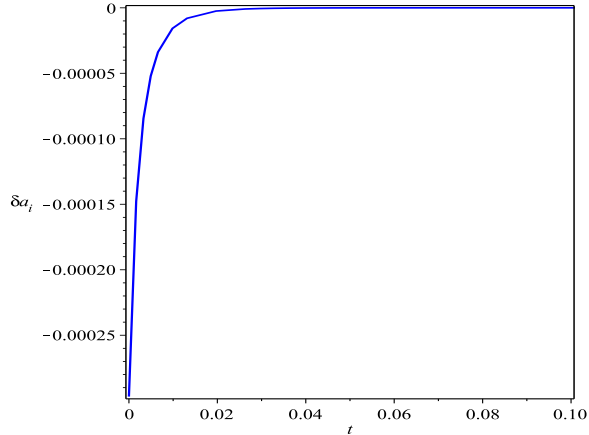}
	\caption{Plot of metric perturbation $\delta a_{i}$ versus $t$. Here $\beta=0.0062, k=0.000016$ and $c=0.$}
\end{figure}
\section{Conclusion}

The present study contributes to the exact solutions of the scalar-tensor theory of gravitation described by S$\acute{a}$ez \& Ballester. 
It is worth mentioned here that the scalar field $\phi$ plays a significant role in the expression for the physical quantities 
$\overline{p},\rho, p ,\xi$ and $\rho_{p}$. We find a point type singularity in the derived models as $p, \rho,\lambda,\rho_{p}$
diverge at $t\to \infty $.\\

The model shows a phase transition from an early decelerating to present the accelerating expansion of the universe. The phase transition 
took place at $z=1.965 \approx 2$. Recently, Hayes {\it et al} \cite{ref65} and Dunlop \cite{ref65} use the comparison of Lyman-$\alpha$ and 
H-$\alpha$ luminosity functions to deduce the range of redshift, which currently is feasible at $z \approx 2$. Thus, $z=1.965$ in 
our derived models is consistent with observational value \cite{ref64,ref65}.\\

Also, our derived models are stable under perturbations. \\

So, we may conclude that our models are improved from earlier works and it presents a better picture of the universe. So it 
deserves attention.\\ 

From Figures $4(a)$, $(b)$ and $(c)$ we observed that the isotropic pressure, in the presence of the bulk viscosity $\lambda$ is a decreasing 
function of time $t$ and approaches to zero at late time but in the absence of bulk viscosity the presence is always negative and tends 
to zero at present time. Thus, we see the role of bulk viscosity for the evolution of the universe.\\

Lastly, we conclude that our derived models deserve attention and show a better shape of the universe.

\section*{Acknowledgments} 
One of the authors (AP) thanks the IUCAA, Pune, India for providing facility and support for a visit where part of 
this work was carried out. 

\end{document}